\documentstyle[epsf,11pt]{article}
\begin{document}
\title{Flux flow resistivity  in the two-gap superconductivity\footnote{To be published in J. Phys. Soc. Jpn.}}
\author{Jun Goryo and Hiroshi Matsukawa \\
{\it Department of Physics and Mathematics, Aoyama Gakuin University}, \\
{\it 5-10-1 Fuchinobe, Sagamihara, Kanagawa, 229-8558, Japan}}
\date{}
\maketitle

\begin{abstract}
We investigate the flux flow state in a two-gap superconductor in which   
two $s$-wave gaps with different amplitudes exist  
on two separate Fermi surfaces. 
The flux flow resistivity is obtained on the basis of the Bardeen-Stephen relation 
and the result agrees well with the anomalous field dependence of the flow resistivity 
recently observed in the two-gap superconductor MgB$_2$. Some typical properties 
of the vortex in this system are also discussed.
\end{abstract}

\begin{flushright}
{\it key words}: MgB$_2$,  two-gap structure, flux flow, extension of Bardeen-Stephen relation
\end{flushright}

The vortex lines in type II superconductors  
are subject to the Lorentz force under an external current and begin to 
flow perpendicular to the current and magnetic field when the Lorentz force 
exeeds the pinning force. 
This is the flux flow state and a finite resistivity arises\cite{Tinkham}.   
The basis of the flux flow resistivity is the presence of 
bound states inside the vortex core. The energy gaps  
of these states are so small that the conductivity in the core is practically 
normal.  Such states were found by Caroli et al. by microscopic methods\cite{CdGM}. 
In both dirty and moderately clean $s$-wave superconductors, 
the flux flow resistivity $\rho_f$ is proportional to the magnetic field $H$\cite{Tinkham}, 
namely, 
\begin{equation}
\rho_f = \frac{H}{H_{c2}} \rho_n,  
\label{ffr-B-S} 
\end{equation}
where $H_{c2}$ is the upper critical field and $\rho_n$ the normal-state 
resistivity. This is the so-called Bardeen-Stephen relation\cite{B-S}. 
In the low-temperature region, this relation holds well for almost the entire field range 
of the vortex state. \cite{matsuda1}

The superconductivity of MgB$_2$ has been investigated with keen interest 
since it has the highest transition temperature ($T_c\simeq39K$) 
among metalic compounds at present and 
a great number of investigations have been carried out\cite{nagamatsu}.  
One of the most characteristic features of this superconductor is that 
two $s$-wave gaps with different amplitudes exist on the two 
separate Fermi surfaces having roughly  equal densities of states (DOSs). 
The two-gap model was proposed on the basis of first-principle calculations,\cite{liu,choi}   
and experimental results 
obtained by point-contact spectroscopy,\cite{point-contact}  
specific heat measurements,\cite{spe-heat} and 
angle-resolved photoemission spectroscopy\cite{arpes} support  the model. 

Recently, the measurement of the flux flow resistivity of 
MgB$_2$ was reported by Shibata et al.\cite{matsuda} 
Large deviation from the $H$-linear dependence for the flux flow 
resistivity has been observed, in spite of the $s$-wave 
pairing symmetry of MgB$_2$\cite{isotope,specific-heat,raman}. 
Such anomalous behavior is expected to be related to the two-gap 
structure, but a clear explanation has not yet been proposed.  
In this study, we investigate the flux flow state in the two-gap system and  
propose a new scenario for the anomalous flux flow resistivity.


First, we investigate the vortex in the two-gap system,  
before discussing the flux flow state. 
Let $\Psi_{\rm L}$ and $\Psi_{\rm S}$ stand for the order parameter for the 
large energy gap and that for the small energy gap, respectively. 
We use a Ginzburg-Landau(GL) free energy 
for the two-gap system in a weak coupling approach with a Josephson-type 
interaction\cite{Zhitomirsky-Dao},   
\begin{eqnarray}
F&=&\int d^3 r \left[ K_{\rm L \perp} |{\bf D}_{\perp} \Psi_{\rm L}({\bf r})|^2 + 
K_{\rm S \perp} |{\bf D}_{\perp} \Psi_{\rm S}({\bf r})|^2 \right. 
\nonumber\\
&&
+ K_{\rm L z} |{\bf D}_{z} \Psi_{\rm L}({\bf r})|^2 + 
K_{\rm S z} |{\bf D}_{z} \Psi_{\rm S}({\bf r})|^2
\nonumber\\
&&
+ \alpha_{\rm L}(T) |\Psi_{\rm L}({\bf r})|^2 +\alpha_{\rm S}(T) |\Psi_{\rm S}({\bf r})|^2
\nonumber\\
&&
 - \gamma \left\{\Psi_{\rm L}^*({\bf r})\Psi_{\rm S}({\bf r}) + \Psi_{\rm L}({\bf r}) \Psi_{\rm S}^*({\bf r})\right\} 
\nonumber\\
&&
\left. +  \frac{\beta_{\rm L}}{2} |\Psi_{\rm L}({\bf r})|^4 + \frac{\beta_{\rm S}}{2} |\Psi_{\rm S}({\bf r})|^4
+\frac{1}{8 \pi} {\bf H}({\bf r})^2\right],  
\label{free-energy}
\end{eqnarray} 
where ${\bf D}={\bf \nabla} - i \frac{2 e}{\hbar c} {\bf A}$ and index $\perp = x,y$. 
We introduce a magnetic field along the $z$-axis.  Let  $\theta_i ({\bf r})$ denote  
the phase of $\Psi_i ({\bf r})$ ($i={\rm L},{\rm S}$). 
There are two types of vortex, one of which 
consists of the winding of $\theta_{\rm L}({\bf r})$ and 
the other consists of that of $\theta_{\rm S}({\bf r})$. 
The temperature dependences of the two gaps\cite{point-contact,arpes} indicate that 
the mixing effect between the large gap and the small gap should play an important role in MgB$_2$.  
This means that the term proportional to $\gamma$ in the GL free energy, eq. (\ref{free-energy}),  
is not negligible.  The term is rewitten as 
\begin{eqnarray}
-\gamma \left\{\Psi_{\rm L}^*({\bf r})\Psi_{\rm S}({\bf r}) + \Psi_{\rm L}({\bf r}) \Psi_{\rm S}^*({\bf r})\right\}=
\nonumber\\
- 2 \gamma |\Psi_{\rm L}({\bf r})||\Psi_{\rm S}({\bf r})|\cos\left(\theta_{\rm L}({\bf r}) - \theta_{\rm S}({\bf r})\right), 
\label{gamma}
\end{eqnarray}
and locks the relative phase as 
\begin{equation}
\theta_{\rm L}({\bf r})-\theta_{\rm S}({\bf r})=
\left\{
\begin{array}{cc}
0 & (\gamma > 0), \\
\pm\pi & (\gamma < 0).  
\end{array}
\right. 
\end{equation}
Therefore, we may state that (i) the two types of vortex have the same $H_{c1}$, and that 
(ii) it is favorable for their cores to be overlapped energetically, i.e., there is an 
attractive interaction between the two types of vortex. 
This phase-locking effect could be modified by some boundary effects\cite{HF} or 
in the thin film system and this possibility will be discussed elsewhere\cite{elsewhere}. 

The coherence length of $\Psi_i({\bf r})$ is\cite{Tinkham}  
\begin{equation} 
\xi_i(T)=\sqrt{-\frac{K_i}{\alpha_i(T)}},
\label{xi}
\end{equation}
and for $|{\bf r}| >> \xi_{\rm L}, \xi_{\rm S}$, the amplitudes of order parameters become constant 
and one obtains a London equation 
\begin{eqnarray}
{\bf \nabla}\times {\bf H} + 
\left(\frac{1}{\lambda^{({\rm L})2}} + \frac{1}{\lambda^{({\rm S})2}} \right)
 \left({\bf A} - \frac{\Phi_0}{2 \pi} \nabla \varphi \right)=0,   
\label{GL-eq}
\end{eqnarray}
where $\Phi_0=h c / 2 e$, 
\begin{equation}
\varphi({\bf r})=\theta_{\rm L}({\bf r})=
\left\{
\begin{array}{ll}
\theta_{\rm S}({\bf r}) & (\gamma > 0), \\
\theta_{\rm S} ({\bf r}) \pm \pi & (\gamma < 0), 
\end{array}
\right. 
\end{equation} 
\begin{equation}
\lambda^{(i)}(T)=\left(\frac{32 \pi e^2 K_{i\perp}}{\hbar^2 c^2}|\Psi_i^{(0)}(T)|^2\right)^{-1/2},  
\label{lambda-i}
\end{equation} 
and $\Psi_i^{(0)}(T)$ is the stationary value of $\Psi_i({\bf r})$ 
in the homogeneous system at a temperature $T$. 
One can see from eqs. (\ref{GL-eq}) and (\ref{lambda-i}) that 
the London penetration depth in this system is    
\begin{equation}
\bar\lambda=\left(\frac{1}{\lambda^{({\rm L})2}}+\frac{1}{\lambda^{({\rm S})2}}\right)^{-1/2}. 
\label{London}
\end{equation}

 
Let us now discuss flux flow resistivity. 
In the two-gap system, the applied current ${\bf J} $ is divided between the band 
with the large gap (L-band) and that with the small gap (S-band);  
 \begin{equation}
{\bf J}={\bf J}_{\rm L} + {\bf J}_{\rm S}. 
\label{ext-j-t-b}
\end{equation}
The divided current  ${\bf J}_i$ provides the Lorentz driving force 
to the vortices in the two bands. 
We assume that the two types of vortex have the same velocity 
because of the presence of the attractive interaction, as we mentioned before. 
The ratio of the two distributed currents is determined to 
minimize the energy dissipation of flux flow, more precisely, the power loss density of the flux flow 
 \begin{equation}
W=W_{\rm L} + W_{\rm S}, 
\label{loss-t-b}
\end{equation}
where 
\begin{equation}
W_i=\pi R^2 \rho_f^{(i)} J_i^2  
\end{equation}
is the power loss per unit cell of the vortex lattice (with a lattice constant $R$) 
per unit length along the $z$-axis\cite{Tinkham} 
and $\rho_f^{(i)}$ is the flux flow resistivity in the $i$-band. The field dependence of $\rho_f^{(i)}$ 
will be discussed later. 
 
To minimize eq. (\ref{loss-t-b}) under the constraint of eq. (\ref{ext-j-t-b}), 
one obtains 
\begin{eqnarray}
{\bf J}_{\rm L}&=&\frac{\rho_f^{({\rm S})}}{\rho_f^{({\rm L})} + \rho_f^{({\rm S})}} {\bf J},  
\nonumber\\
{\bf J}_{\rm S}&=&\frac{\rho_f^{({\rm L})}}{\rho_f^{({\rm L})} + \rho_f^{({\rm S})}} {\bf J}. 
\label{div-cur}
\end{eqnarray}
Then, 
\begin{equation}
W=\pi R^2 \frac{1}{1/\rho_f^{({\rm L})}+ 1/\rho_f^{({\rm S})}} J^2,  
\end{equation}
and this equation indicates that the total flux flow resistivity 
in this system is 
\begin{equation}
\rho_f^{two}=\frac{1}{1/\rho_f^{({\rm L})} + 1/\rho_f^{({\rm S})}}, 
\label{rho-f-two}
\end{equation}
{\it which is the parallel connection of the resistivity in the L-band  
and that in the S-band.} 

Let us discuss the field dependence of $\rho_f^{(i)}$. Since the two gaps have 
the $s$-wave pairing symmetry, 
these two resistivities are considered to obey the Bardeen-Stephen relation, i.e., 
\begin{equation}
\rho_f^{(i)}=\frac{H}{H_{c2}^{(i)}} \rho_n^{(i)} 
\label{ffr-1-2}
\end{equation}
for $H_{c1} < H <H_{c2}^{(i)}$, where $\rho_n^{(i)}$ is the normal-state resistivity in  
 the $i$-band, and 
 \begin{equation}
 H_{c2}^{(i)}=\frac{\Phi_0}{2 \pi \xi_i^2}.  
 \label{hc2}
 \end{equation}
 For MgB$_2$, the results of the point-contact spectroscopy experiment suggest that 
 the small gap is suppressed quicker than the large gap.\cite{point-contact} This implies 
 that $\xi_{\rm L} < \xi_{\rm S}$ and $H_{c2}^{(\rm S)}<H_{c2}^{(\rm L)}$. 
The value $H_{c2}^{(\rm L)}$  coincides with the upper critical field in this system. 
The kink point that has been observed in the field dependence 
of the specific heat\cite{spe-heat} corresponds approximately to $H_{c2}^{(\rm S)}$,  
since the presence of the kink indicates that the small-gap pairing is almost suppressed\cite{note}. 
Then, the S-band would quite close to the normal metalic state 
above $H_{c2}^{(\rm S)}$. This picture is also supported by the 
calculation for quasiparticle DOS in the vortex state.\cite{Ichioka,Kusunose} 
Therefore, it is possible to approximate  
 \begin{eqnarray}
 \rho_f^{(\rm S)}(H) \simeq \rho_n^{(\rm S)}  
 \label{rho-n}
 \end{eqnarray}
 for $H_{c2}^{(\rm S)}<H<H_{c2}^{(\rm L)}$. 
 

Then, we obtain the field dependence of $\rho_f^{two}$ in eq. (\ref{rho-f-two}). 
By using  eqs. (\ref{ffr-1-2}) and (\ref{rho-n}), 
\begin{equation}
\rho_f^{two}
=
\left\{
\begin{array}{ll}
\frac{H}{H_{c2}^{(\rm L)}/\rho_n^{(\rm L)}  + H_{c2}^{(\rm S)}/\rho_n^{(\rm S)}} 
& (H_{c1}<H \leq H_{c2}^{(\rm S)}),\\
& \\
\frac{H}{H_{c2}^{(\rm L)}/\rho_n^{(\rm L)}  + H/\rho_n^{(\rm S)}} & 
(H_{c2}^{(\rm S)} < H \leq  H_{c2}^{(\rm L)}).  
\end{array}
\right.
\label{field-vs-rho-f-two} 
\end{equation}
We can see that $\rho_f^{two}$ is linear with $H$ in the low-field regime, 
curves convexly in the high field regime and is continuously connected to 
the normal-state resistivity in the two-band system $(1/\rho_n^{(\rm L)} + 1/\rho_n^{(\rm S)})^{-1}$.  

\begin{figure}
\begin{center}
\leavevmode
\epsfxsize = 10.0cm
\epsfbox{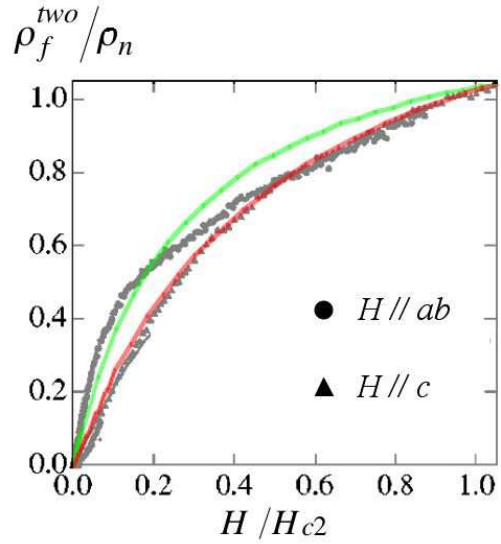}\\
\vspace*{0cm}
\caption{We compare the experimental 
data of flux flow resistivity in MgB$_2$ given in ref. 11 and 
our results calculated with eq. (\ref{field-vs-rho-f-two}). 
Black dots and triangles denote experimental results for the $H \parallel ab$ case 
and $H \parallel c$ case, respectively. The green line
denotes eq. (\ref{field-vs-rho-f-two})  with parameters 
$H_{c2}^{(\rm S)} / H_{c2}^{(\rm L)} = 0.02$ and $\rho_n^{(\rm S)}/\rho_n^{(\rm L)}=0.26$, and  
the red line denotes eq. (19) with $H_{c2}^{(\rm S)} / H_{c2}^{(\rm L)} =0.1$ 
and $\rho_n^{(\rm S)}/\rho_n^{(\rm L)}=0.5$. The red line agrees very well with the data for $H \parallel c$.} 
\label{fig1}
\end{center}
\end{figure}

Let us compare eq. (\ref{field-vs-rho-f-two}) to the experimental results of the flux flow resistivity 
in MgB$_2$\cite{matsuda}. The measurements were performed with 
the magnetic field parallel to the $c$-axis of the crystal and also with the field in the $ab$-plane\cite{matsuda}. 
The kink point that has been observed in the field dependence of 
the specific heat\cite{spe-heat} suggests that $H_{c2}^{(\rm S)} / H_{c2}^{(\rm L)} = 0.02$ 
for the $H \parallel ab$ case and $H_{c2}^{(\rm S)} / H_{c2}^{(\rm L)} =0.1$ for the $H \parallel c$ case\cite{comment}. 
Unfortunately, there is no definite information for the ratio $\rho_n^{(\rm S)}/\rho_n^{(\rm L)}$ at 
present, 
and  we use it as a fitting parameter and choose 
$\rho_n^{(\rm S)}/\rho_n^{(\rm L)}=0.26$ for the $H \parallel ab$ case 
and $\rho_n^{(\rm S)}/\rho_n^{(\rm L)}=0.5$ for the $H \parallel c$ case.
As is shown in Fig. \ref{fig1}, it should be emphasized that
 both cases of the experimental results\cite{matsuda} are consistent with the convex-type behavior 
 of eq. (\ref{field-vs-rho-f-two} ), particularly for the $H \parallel c$ case. 
The robustness of the coincidence in the $H \parallel c$ case is supported 
by the fact that the  20$\%$ change in the parameter $\rho_n^{(\rm S)}/\rho_n^{(\rm L)}$ around the value 0.5 causes at most a 10$\%$ change in $\rho_f^{two}$. The difference between our 
calculated results and the experimental results in the high-field region of the $H \parallel ab$ case is considered to be related to complications of the analysis of experimental data, since the vortices driven by the microwave 
field\cite{matsuda} move in the ac-plane and the anisotropy of the crystal causes the current-direction dependence of the flux flow resistivity. The experimental data for the $H \parallel ab$ case correspond to the  averaged value of this anisotropy.

\vspace*{0cm}



In summary, we have considered a two-gap superconductor such as MgB$_2$.\cite{nagamatsu}
In this system, there are two bands that cross 
the Fermi level in the metalic phase and 
two $s$-wave gaps with different amplitudes   
that arise on these two Fermi surfaces.  
There are two types of vortex, one of which consists of the 
phase winding of the order parameter for the large gap and 
the other consists of that for the small gap. 
These two types of vortex have the same lower critical field and 
attract each other. 
We have examined the flux flow resistivity in this system. 
The two types of vortex have the same velocity because of 
the presence of the attractive interaction. The flux flow resistivity 
in the two bands is considered to obey the Bardeen-Stephen relation,  
since the two gaps have the $s$-wave pairing symmetry.   
We have obtained that, to minimize the power loss (energy dissipation per unit time) caused by 
the flux flow, {\it the total flux flow resistivity 
is expressed as a parallel connection of the resistivity in the band with the large gap and 
that in the band with the small gap.} It should be noted that 
the amplitude of the small gap is suppressed faster  
than that of the large gap by the magnetic field in MgB$_2$\cite{point-contact}, 
and the resistivity in the band with the small gap 
is approximately equal to its nomal state value in the field 
region where the small gap is almost entirely suppressed.   
Taking this into account, our formula agrees quite well with the anomalous field dependence of 
the flux flow resistivity as observed experimentally\cite{matsuda}.

The authors are grateful to J. Akimitsu, H. Fukuyama, N. Furukawa, K. Kubo, K. Ishikawa, K. Izawa, 
Y. Matsuda, S. Soma, P. Thalmeier, and K. Maki for useful discussions  
and comments.  This work is financially supported by Grant-in-Aid for Scientific Research from Japan Society for the Promotion of Science under Grant No. 15540370 and No. 16740226.


\begin{thebibliography}{99}
\bibitem{comment} Another crossover magnetic field is reported 
by ref. 11, but its origin is not clear at present.
\end{thebibliography}
\end{document}